%% file: lago-eosc-icrc2021.tex
\documentclass[a4paper,11pt]{article}
\usepackage{pos}
\usepackage{amsfonts}
\usepackage{amsbsy}
\usepackage{multirow}

\title{The EOSC-Synergy cloud services implementation for the Latin American Giant Observatory (LAGO)}
\ShortTitle{LAGO Thematic Service at EOSC-Synergy}

\author*[a]{A.J. Rubio-Montero}
\author[a]{R. Pagán-Muñoz}
\author[a]{R. Mayo-García}
\author[b]{A. Pardo-Diaz}
\author[c]{I. Sidelnik}
\author*[d]{H. Asorey}

\affiliation[a]{Centro de Investigaciones Energéticas, Medioambientales y Tecnológicas (CIEMAT)\\
Av. Complutense 40, 28040 Madrid, Spain}

\affiliation[b]{Centro Extremeño de Tecnologías Avanzadas (CETA-CIEMAT)\\
Calle Sola 1, 10200 Trujillo, Spain}

\affiliation[c]{Instituto de Tecnologías en Detección y Astropartículas (ITeDA, CNEA/CONICET/UNSAM)\\
Centro Atómico Constituyentes, Av. General Paz 1499, 1450 Villa Maipú, Buenos Aires, Argentina}

\affiliation[d]{Departamento de Física de Neutrones, Centro Atómico Bariloche (CNEA/CONICET)\\
Av. Bustillo 9500, 8400 San Carlos de Bariloche, Argentina}

\forColl{LAGO} 

\emailAdd{hernan.asorey@iteda.cnea.gov.ar}

\abstract{The Latin American Giant Observatory (LAGO) is a distributed cosmic ray observatory at a regional scale in Latin America, by deploying a large network of Water Cherenkov detectors (WCD) and other astroparticle detectors in a wide range of latitudes from Antarctica to México, and altitudes from sea level to more than 5500 m a.s.l. Detectors telemetry, atmospherics conditions and flux of secondary particles at the ground are measured with extreme detail at each LAGO site by using our own-designed hardware and firmware (ACQUA).
To combine and analyse all these huge amount of data, LAGO developed ANNA, our data analysis framework. Additionally, ARTI, a complete framework of simulations designed to simulate the expected signals at our detectors coming from primary cosmic rays entering the Earth atmosphere, allowing a precise characterization of the sites in realistic atmospheric, geomagnetic and detector conditions.

As the measured and synthetic data started to flow, we are facing challenging scenarios given the large amount of data emerging, performed on a diversity of detectors and computing architectures and e-infrastructures. These data need to be transferred, analyzed, catalogued, preserved, and provided for internal and public access and data-mining under an open e-science environment. In this work, we present the implementation of ARTI at the EOSC-Synergy cloud-based services as the first example of LAGO' frameworks that will follow the FAIR principles for provenance, data-curation and re-using of data.

For this we calculate the flux of secondary particles expected in up to 1 week at detector level for all the 26 LAGO, and the 1-year flux of high energy secondaries ($p_S>800$\,GeV/c) expected at the ANDES Underground Laboratory and other sites. Therefore, we show how this development can help not only to LAGO but other data-intensive cosmic rays observatories, muography experiments and underground laboratories.}

\FullConference{37$^{\textrm{th}}$ International Cosmic Ray Conference (ICRC 2021)\\
		July 12th -- 23rd, 2021\\
		Online -- Berlin, Germany}

\begin{document}
\maketitle

\section{Introduction}\label{sec:intro}

Astroparticle physics is nowadays one of the scientific fields that evidence large interdisciplinary contributions due to its large variety of topics, e.g., high energy astrophysics, detector physics and computational sciences.
The Latin American Giant Observatory, managed by the LAGO Collaboration\footnote{A cooperative and non-centralized collaboration of more than 30 institutions from 10 Iberoamerican countries}, is an extended astroparticle observatory providing key information about space weather and climate-related phenomena, performing scientific research on background radiation detection at ground level.
Thanks to the development of several computational tools, it is possible now to simulate the complete response of the system, but standardized computational mechanisms are needed to profit from the simulation framework\,\cite{Asorey2018, Sarmiento2019}, mainly on the coordination, share and curation of the results.
The European Commission (EC) is fostering initiatives in this direction that being able to guarantee the long-term sustainability of the research.
For instance, by adopting the open science paradigm with the FAIR (Findable, Accessible, Interoperable, Re-usable data) principles, and the federation of virtual resources and organisations.
Consequently, improved simulations methods and results are presented in this work within the EOSC-Synergy project.

\input{lago-eosc-icrc2021-lago}
\input{lago-eosc-icrc2021-computations}
\input{lago-eosc-icrc2021-results}

\section{Conclusions and future work}\label{sec:conclusions}

The evolution of high energy and astroparticle physics demands an increasing amount of computational resources to get more accurate results. 
While the computing and storage capabilities continue growing, 
there is a pressing need to standardize the synthetic and measured data production, analysis, curation and access.
For LAGO, the integration of EOSC ecosystem and the standardization of the computations are focused on the definition, generation and storage of the produced data and metadata by tools deployed in Docker images.


In this work, we show the first results of how ARTI has been successfully adapted to the cloud environment, and thus, by increasing integration times and energy ranges for the calculation of the expected flux of particles at all the LAGO sites.
Even more, for some planned uses of this framework, such as the study of high energetic secondary particles at the tens of TeV scale for underground laboratories background calculations and expected flux of muons for muography geophysical applications, the first results were shown for the future ANDES, where we obtained for the first time the expected flux over an integration time of 1 year at different altitudes of the ANDES mountain profile.
In all those cases, the accuracy and statistical significance of previous results have been extended by using these new resources.

\subsection*{Acknowledgements}

The LAGO Collaboration is very thankful to the Pierre Auger Collaboration for their continuous support.
It is also acknowledged the support of the LAGO Collaboration members.
This work has been partially carried out on the ITeDA cluster, we thank A.P.J. Sedosky-Croce for his continuous support.
This work was carried out within the 'European Open Science Cloud - Expanding Capacities by building Capabilities' (EOSC-SYNERGY) project, founded by the European Commission’ Horizon 2020 RI Programme under Grant Agreement nº 857647.
HA thanks to R. Mayo-García and CIEMAT for their warm welcome and support at CIEMAT in Madrid.

\footnotesize
\bibliographystyle{abbrv}
\bibliography{lago-eosc-icrc2021}

\input{author}
%
%

\end{document}

%% file: lago-eosc-icrc2021-lago.tex
\section{The Latin American Giant Observatory (LAGO)}\label{sec:sc_lago}

The LAGO detection network consists of single or small arrays of WCD installed in different sites across the Andean region.
As shown in the right panel of Figure~\ref{fig:figlagoarch}, this network spans a region from the south of Mexico, with a small array installed at $4,550$\,m a.s.l.\ in Sierra Negra (México), to the Antarctic Peninsula.
LAGO is distributed over similar geographical longitudes covering also a wide range of geographical latitudes (thus sweeping very different geomagnetic rigidity cutoffs) and altitudes (different atmospheric absorption depths).


The LAGO programs cover all the aspects of the project, from the design and deployment of new detectors, the calibration and operation of the detection network, the data analysis, simulation and interpretation, and even finding new ways to transfer data from remote sites and to handle the data produced.
By performing precise measurements of the time evolution of the atmospheric radiation at ground level, we are focused on searching for high energy transient phenomena and in the monitoring of space weather long-term and transient phenomena.



By harnessing the advantages of the WCDs, it is possible to determine the flux of secondary particles at different bands of deposited energy within the detector volume.
These bands are dominated by different components of the secondary particles produced during the interaction of GCR with the Earth's atmosphere.
This multi-spectral analysis technique (MSAT) constitutes the basis of our Space Weather oriented data analysis. 
By combining all the data measured at different locations of the detection network, the LAGO project is able to provide information on the temporal evolution of the disturbances produced by different transient and long-term space weather phenomena, some of them even on an almost real-time basis.


Complete simulation sequences involve all the related aspects of our objectives: from CR propagation to detector response\,\cite{Asorey2018,Sarmiento2019}; the data analysis techniques specially designed for the very different energy and temporal scales of the studying phenomena; 
to the novel implementation of cloud-based synthetic data production and data curation\,\cite{RubioMontero2021}, shown in this work. 

For this purpose we developed ARTI, a framework of interacting codes focused on estimating the cosmic ground radiation flux and the WCD's response to it, by joining:
the Global Data Assimilation System (GDAS) to obtain the physical atmospheric profiles\,\cite{Grisales2021}; MAGCOS, an application to simulate the evolving conditions of the geomagnetic field;
CORSIKA, a complete Monte Carlo code to simulate the interaction of astroparticles with the atmosphere;
and, finally, the Geant4 toolkit to obtain the expected response of the LAGO detectors\,\cite{Agostinelli2003}.

The first step is to calculate the expected number of primaries by integrating the measured flux  ($\Phi$) of each GCR species ($1<Z<26$) within the $1<E_p/\mathrm{GeV}<10^6$ energy range in $1$\,m$^2$ of atmosphere at an altitude of $112$\,km a.s.l.
For the geomagnetic impact calculation, the 13$^\mathrm{th}$ generation of the International Geomagnetic Reference Field IGRF13-2019, and the Tsynganekov (TSY2001) models are taking into account as described in~\cite{Asorey2018}.
 Here, $\Phi$ is obtained from the integral of the spectrum of each type of primary reaching the Earth's atmosphere for the whole sky hemisphere ($0\leq \theta \leq \pi/2$, $-\pi \leq \varphi \leq \pi$) above the detector.
For a $1$\,m$^2$ detector and an integration time of one hour, a typical run per site may require the simulation of $\sim (2-5)\times 10^7$ primaries, depending on the minimum rigidity cutoff of the site ($\lesssim 2$\,GV for the LAGO antarctic sites).

The local atmospheric response to the primary flux is obtained by using CORSIKA\,\cite{Corsika}.
Mainly, we extract the corresponding atmospheric profile of the site from GDAS, to obtain a better estimation of the flux of secondaries at the detector level, as large differences have been observed when compared with the more general MODTRAN profiles\,\cite{Grisales2021}. 
Geomagnetic effects are taken into account by constructing a time-dependent tensor containing the directional rigidity cutoff of the site and including the geomagnetic field disturbances. 
Only those secondaries originated by primaries with directional rigidities above the corresponding tensor values are included in the calculation.
This method has the advantage of allowing the usage of the same site simulation set in the evolving conditions of the geomagnetic field.
As the tensor depends on time, we are also able to calculate the expected time-dependent flux at the ground in any place of the World.
Furthermore, a method was developed to deal with those incoming particles traversing the penumbra\,\cite{Asorey2018}.

Finally, only the selected secondary particles are injected into the Geant4-based simulated detectors\,\cite{Sarmiento2019}.
The model considers the detector geometry and materials, the inner coating, the PMT model (characterized by its geometry, quantum efficiency to different photon wavelengths, electronics and time response) and even the water quality and the presence of different solutes.
The Cherenkov photons produced by the secondary particles in the water volume are propagated through the detector and finally, the expected pulses are obtained and stored.
The final results are the time series of the expected flux of secondaries, and the simulated charge histograms that contain information about the deposited energy in the detector.
These figures are important for the detector calibration, as the deposited charge in the PMT for a central and vertical muon (VEM) going through the detector is the main calibration parameter and depends on factors, such as the water and coating quality, the detector geometry, the detector trigger or the PMT polarization voltage.


Therefore, ARTI enables to automatically determine accurately the time-dependent flux of signals expected at detectors of any type, in any site around the World, and under realistic atmospheric and geomagnetic time-evolving conditions.
ARTI data-sets are catalogued following a hierarchical and pipelined scheme:
    {\textbf{S0}}: CORSIKA secondary outputs;
    {\textbf{S1}}: analysis and outputs of the S0 data-set (expected flux of secondary particles and the corresponding parent primary); and
    {\textbf{S2}}: detector response simulation (time series of signals in the detector and the calibration charge histograms).



%% file: lago-eosc-icrc2021-computations.tex
\section{Standardizing LAGO Computations}\label{sec:standardising-lago-computations}

Unlike other observatories, LAGO has not a unique location.
Instead, detectors are distributed at a continental scale and are interoperated by a consortium integrated by all the participant institutions.
Therefore, several challenges appear for the development, management and sharing of resources and codes. There is a lack of unified computational and storage resources, common accessibility methods, and effective mechanisms for the curation and sharing of the generated data.

Clearly, the complexity and size of LAGO made necessary a collaborative approach, which can be framed within the open science paradigm. This will facilitate a more transparent research, the accessibility to the resources, and it will help to disseminate and re-use the knowledge obtained.
To assure the continuity of research, the European Commission (EC), one of the major scientific LA's partners, promotes open science and force to standardize the scientific methods and processes within its funded projects.
Two are the main objectives for the LAGO case: to migrate into open data through the adoption of the FAIR principles for the generation of measured and synthetic data, as well as in their storage and publication; and to make use of the services and mechanisms available at the European Open Science Cloud (EOSC), a trusted virtual environment for hosting and processing research data to support the EC science.

LAGO has already integrated several services from the EOSC marketplace through its standardization. 
To enable the federation of identities, specifically for research communities as virtual organisations (VOs), EduTeams was used.
This is an identity provider (IdP) supported by GEANT, and allows delegating credentials to other IdPs such as EGI Check-in, which is the prevalent in EOSC. However, it is not tied to that infrastructure, delegation can benefit from other public clouds such as AWS, Google or Azure, or even local HPC facilities.
Following the EOSC adoption, providers from EGI FedCloud are preferentially used.
As every provider offers its resources ``as they are'', researchers should make use of other utilities for building pre-configured virtual clusters, one per provider with the IM (Infrastructure Manager) service, or deploying cross-provider infrastructures with the EC3 (Elastic Cloud Computing Cluster) service.

On the other hand, we selected the EGI DataHub to store whole data and metadata generated.
Based on OneData, a globally distributed cloud filesystem, it can be easily deployed in several LA partners, potentially growing the capacity to exabytes.
In this sense, the metadata was standardized following JSON-LD 1.1 (W3C), DCAT2 (W3C) and DCAT-AP2 (EC) schemas.
Additionally, persistent identifiers (PiDs) for every data catalogue are obtained from Handle.net trough the B2HANDLE service before their publication and, B2FIND, a customised CKAN repository, can harvest metadata from OneData providing a simple way to find and access data collections.



\begin{figure}[!ht]
  \centering
  \includegraphics[scale=0.55]{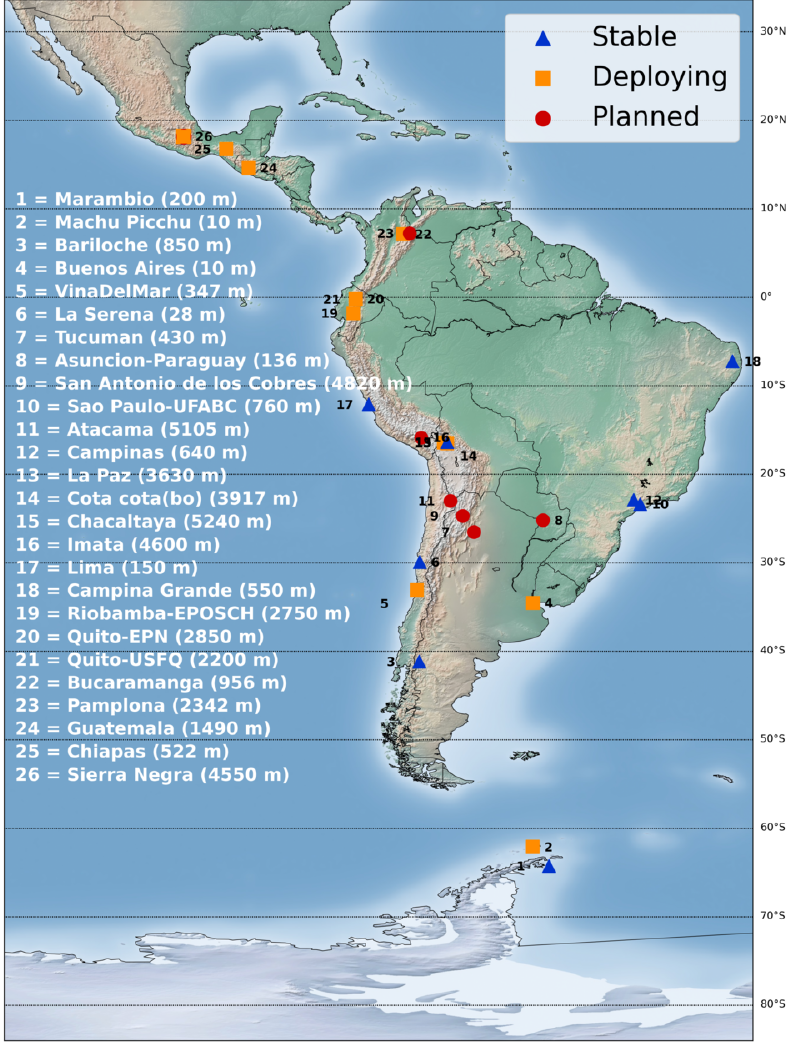}
  \includegraphics[scale=0.23]{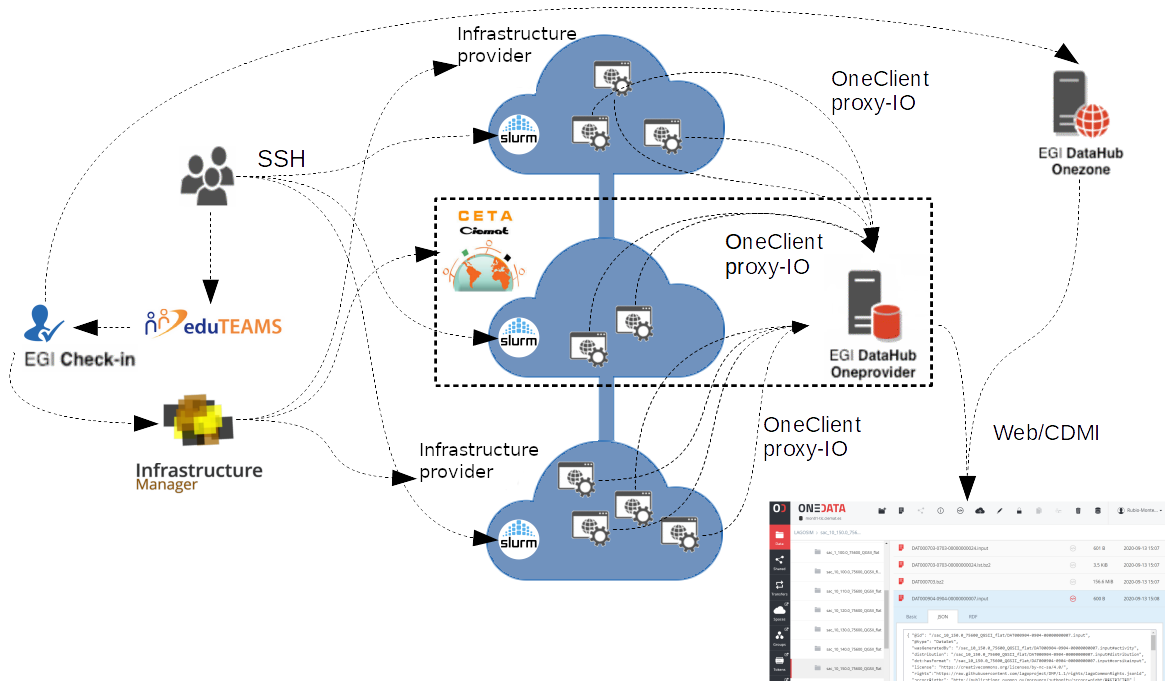}
  \caption{(left) The Latin American Giant Observatory water Cherenkov network detection: operational detectors (blue triangles), under development (orange squares), and planned sites (red circles). (right) The architecture of the LAGO computation showing the large variety of used resources and the complexity of the cloud based implementation of the LAGO simulation services.}
  \label{fig:figlagoarch}
\end{figure}

As described in Rubio-Montero et al.\,\cite{RubioMontero2021}, the architecture follows a basic design: the core intelligence of simulations and data analysis are packed in Docker images, being able, after deployment in a virtual cluster, to automatically produce, check, store and publish the results in DataHub, with the adequate metadata to be referenced by PiDs and used by harvesters at a later stage. 
As the whole computation is self-contained in the Docker image, the production can be easily performed on virtual infrastructures deployed by services such as EC3/IM on cloud resources, or even manually in private clusters.

From the user’s point of view, the architecture of this implementation is shown in Figure~\ref{fig:figlagoarch}.
Depending on the task, the corresponding dockers are installed and run, and data and metadata can be retrieved or stored in the DataHub repository in the cloud.
Authorized users can download, analyse, create or even transform old data.
However, uploading data to the official repository in DataHub is restricted only when it is done via the official versions of the LAGO software, only available as certified Docker images.
Currently, the developed dockers are the S0 ARTI-Sim docker, containing the application that encapsulates CORSIKA v7.7402 binaries and LAGO-ARTI simulation tools;
the S1 ARTI-Analysis docker, containing the LAGO-ARTI analysis tools;
and the S2 ARTI-Detector docker, including ROOT, Geant4 and the LAGO-ARTI detector simulation.

%% file: lago-eosc-icrc2021-results.tex

\section{Capabilities and first results} \label{sec:capabilities-and-results}
As part of the demonstration stage of the LAGO thematic service in EOSC-Synergy, we deployed a cloud-based production of synthetic data within the EOSC services, involving the usage of a constrained set of cloud resources provisioned by EOSC and the S0 (ARTI-Sims) and S1 (ARTI-Analysis) dockers.
By using the EOSC services described in the previous section, we set up a virtual cluster based on a Slurm Workload Manager job scheduler with one virtual master (v-master) and 10 v-nodes with 16 Intel Xeon Core E7 processors and 250 GB of shared memory and disk each one (Docker can use both unified or equally).

The number of injected primaries per site was calculated as described in section\,\ref{sec:sc_lago} and are based on previously published works\,\cite{Asorey2018, Sarmiento2019}, for a detector of $1$\,m$^2$, for protons to irons ($1\leq Z \leq 26$), where $Z$ is the primary atomic number, and by setting the time integration parameter $t$ from 1 day to 1 year depending on the studied case.

For each geographical site, the flux was calculated by using the yearly-averaged GDAS local atmosphere\,\cite{Grisales2021}, and assuming an isotropic flux of primaries for a full hemisphere sky, i.e., ($0\leq\theta\leq\pi/2$, $-\pi\leq\varphi\leq\pi$), impinging on a volumetric (non-flat) detector for the energy range $\min(R_C(\phi,\lambda,\theta,\varphi,t_0)) \times Z \leq E_p / \mathrm{GeV} \leq 10^6$, where $(\theta,\varphi)$ are the primary arrival direction, $E_p$ is the primary energy, and $R_C$ is the local rigidity tensor, depending on the primary trajectory $(\theta,\varphi)$, the geographic coordinates of the site $(\phi,\lambda)$ and on the current time $t_0$.
Given the hard primary spectrum $j(E,Z)=j_0(Z) E^{-\alpha(E,Z)}$, with typical values for the spectral index $\alpha(E,Z)$ varying between $-2.7$ and $-3.1$ in this energy range, the total number of primaries strongly depends on $\min(R_C)$, and thus, we considered the minimum of the rigidity tensor for the S0 calculation stage, and after that, during the S1 analysis stage, we selected only those secondaries originated by primaries in allowed trajectories, included the corresponding treatment of the penumbra trajectories\,\cite{Asorey2018}.

During this first test run, we simulate the expected integrated flux at each of the current and planned LAGO sites for a minimum integration time of $1$\,day and up to $7$\,days of flux for high altitude or high latitude sites.
Given the stochastic nature of these calculations, as the integration of the flux is increased, the impact of the statistical fluctuations is reduced, and correspondingly, it is enlarged their statistical significance and their physical relevance.
Moreover, at some sites like Tucumán (TUC, Argentina), San Antonio de los Cobres (SAC, Argentina) or Quito (QUI, Ecuador), we calculated the expected flux at different altitudes bearing in mind the possibility to install several detectors in the same site at different altitudes.

\begin{figure}[!ht]
  \centering
  \includegraphics[width=0.329\textwidth]{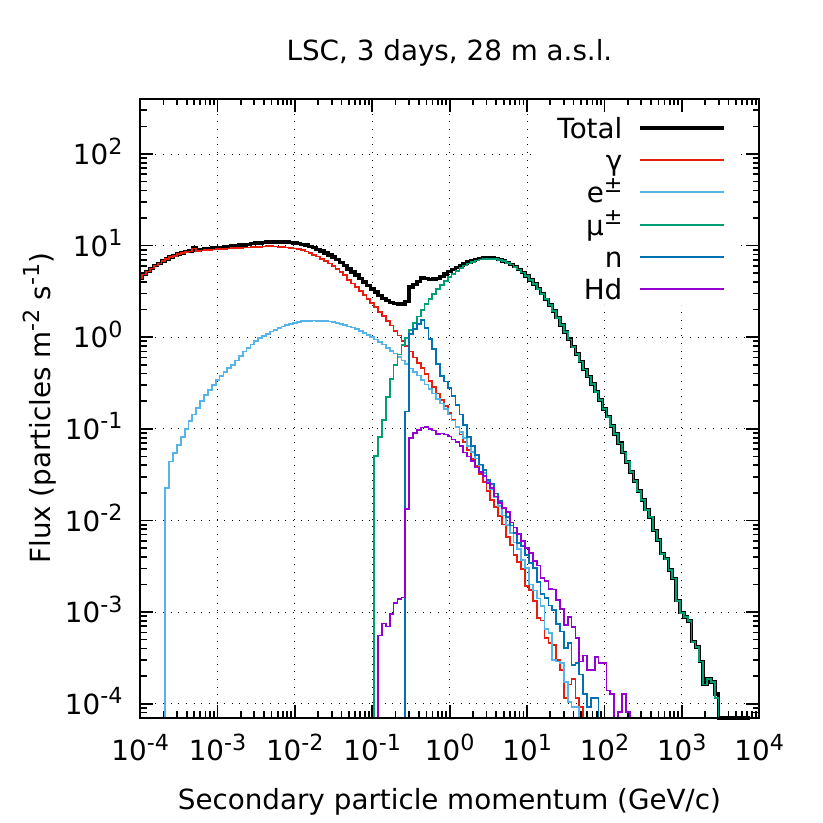}
  \includegraphics[width=0.329\textwidth]{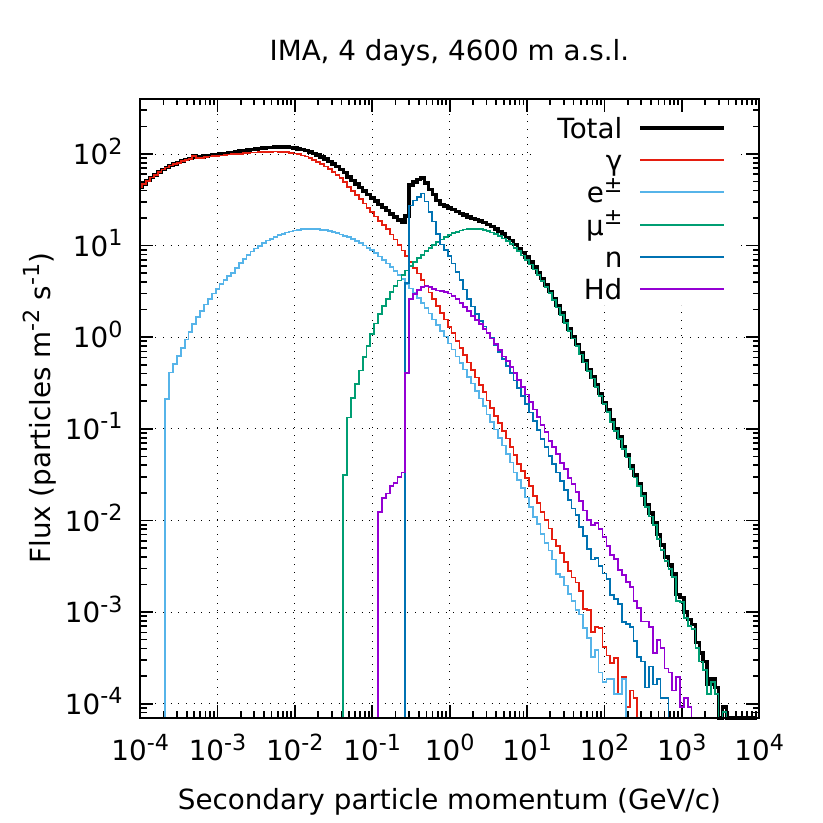}
  \includegraphics[width=0.329\textwidth]{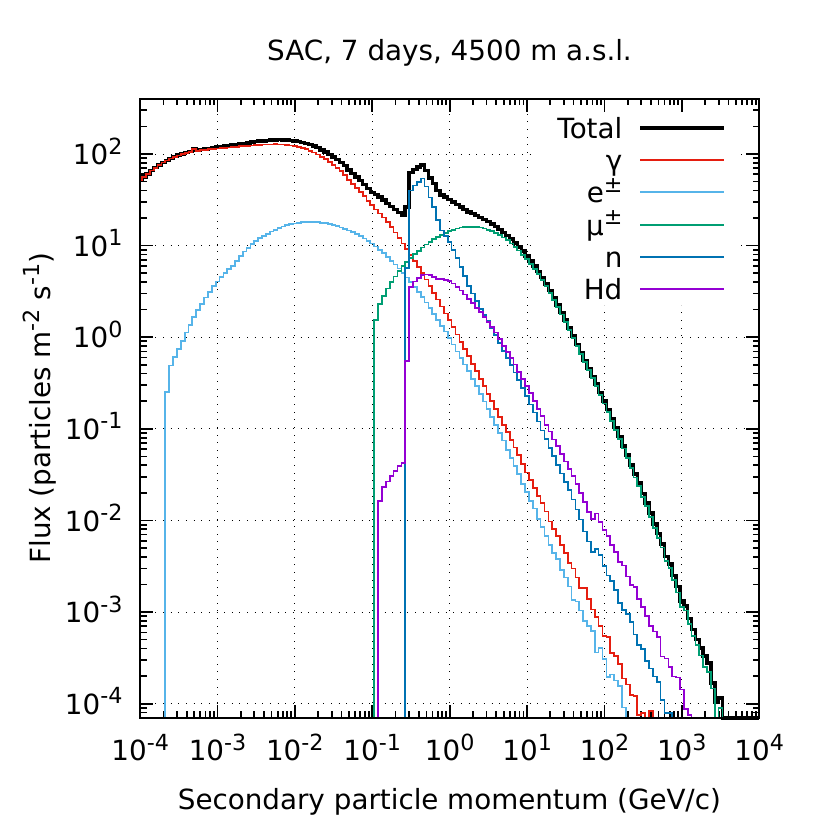}
  \caption{The energy spectrum of the flux of secondary particles expected at three LAGO sites, LSC (La Serena, Chile, sea level, left);  IMA (Imata, Perú, at $4600$\,m a.s.l., centre), and SAC (San Antonio de los Cobres, Argentina, at $4500$\,m a.s.l., right). The atmospheric absorption is clearly visible.}
  \label{fig:secondaries}
\end{figure}

In Figure~\ref{fig:secondaries} it can be seen the secondary spectra for three LAGO sites: La Serena (LSC), Chile, at sea level (left); Imata (IMA), Perú, $4600$\,m a.s.l. (centre), and San Antonio de los Cobres (SAC), Argentina, $4500$\,m a.s.l. (right).
Results from San Antonio de los Cobres at non-standard altitudes will be shown in a dedicated study\,\cite{sarmiento2021}.
Atmospheric absorption at sea level detectors is clearly visible in these plots.
These spectra are used to evaluate the expected flux of secondaries and determine, e.g., the optimal detector geometry, the detector calibration parameters, or the typical energies expected for the different components of the atmospheric radiation background.
Even more, the presence of the neutron peak in the flux motivate the recent evaluation of increasing the neutron sensitivity for space weather studies by doping the detector's water with NaCl\,\cite{Sidelnik2020}.

Additionally, as high energy muons are the main source of signals for muography studies\,\cite{Calderon2021}, and of background radiation at underground laboratories, we extended the calculations to those non-LAGO sites of interest in several studies we are carrying out.
In particular, we estimate the expected flux at potential muography candidate sites in Latin America and for several underground laboratories around the World, with special emphasis on the projected site of the future ANDES Underground Laboratory.

\begin{figure}[!ht]
  \centering
  \includegraphics[width=0.32\textwidth]{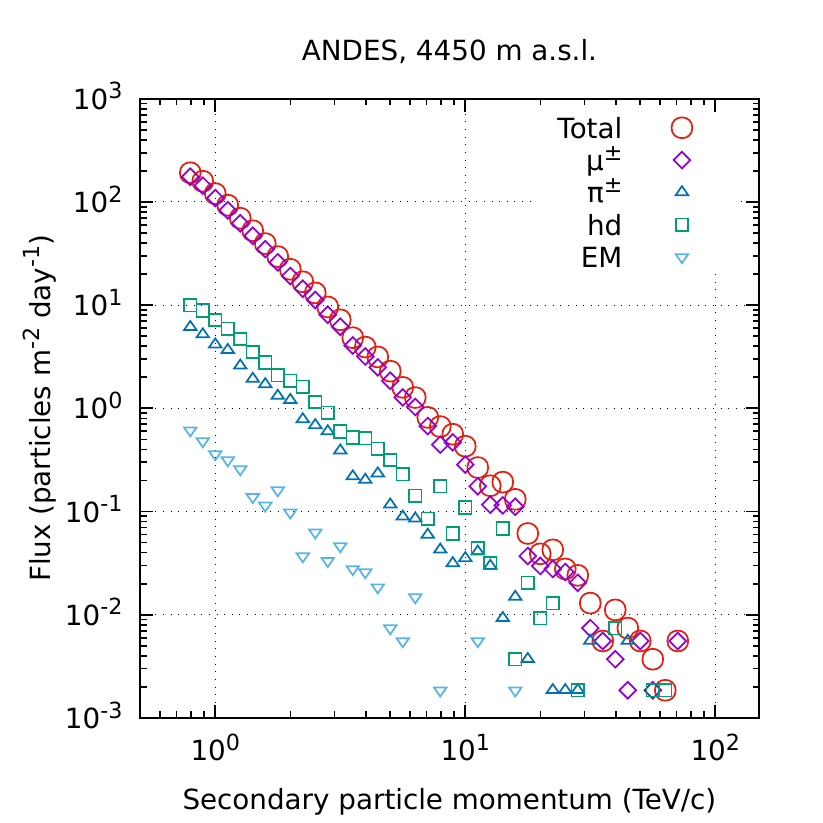}
  \includegraphics[width=0.32\textwidth]{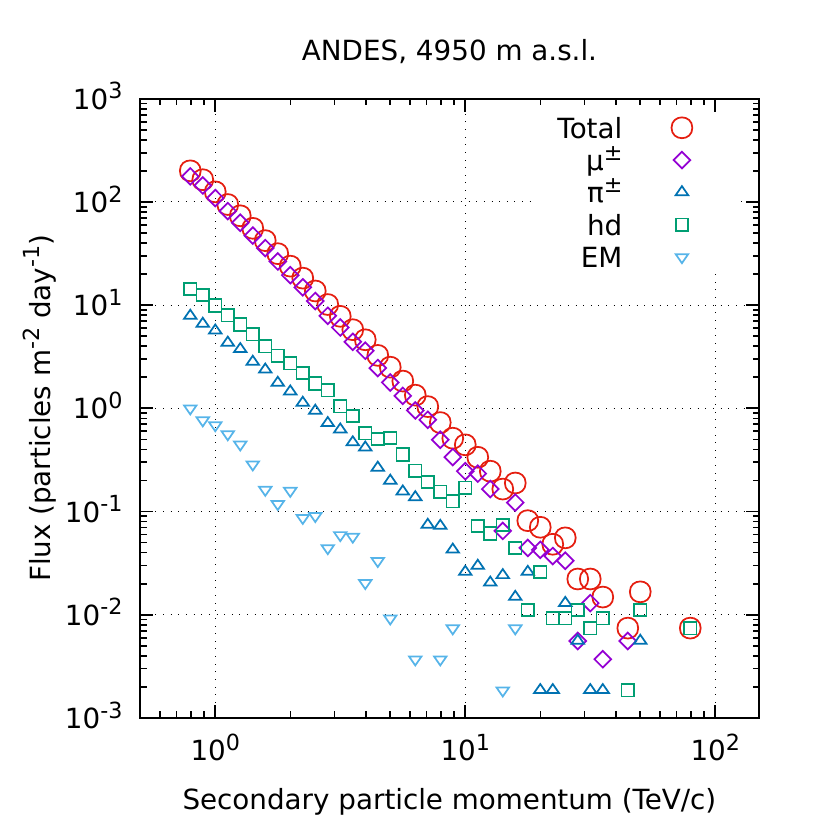}
  \includegraphics[width=0.32\textwidth]{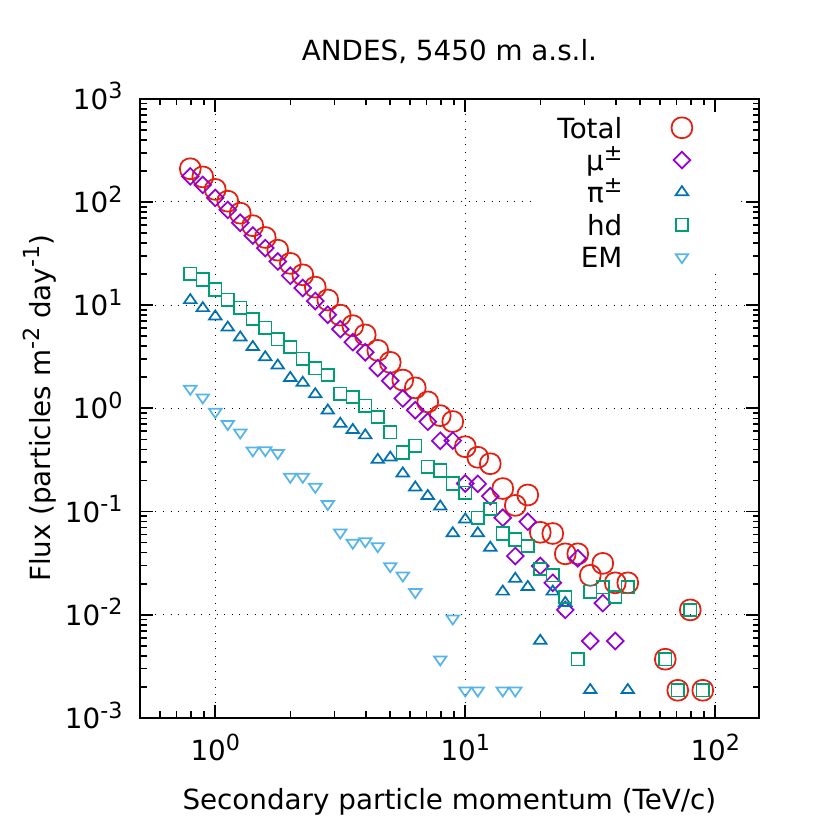}
  \caption{From left to right, the normalized expected flux of secondaries at the TeV/c scale reaching the ANDES mountain at $4450$, $4950$ and $5450$\,m a.s.l., corresponding to an integration time of 1 year at each site. Please notice at the higher energy not only muons but charged pions and nucleons are present. 
  }
  \label{fig:he}
\end{figure}

As in these cases, we need to consider only the higher energy component of the secondary flux at ground, we selected only those secondaries with energy $E_S > 800$\,GeV, as only those particles could be able to penetrate hundreds to thousands meters of rock.
By doing these we were able also to increase the integration time $t$ of the flux, and for the case of, e.g., ANDES we reached $t=1$\,year.
In all cases, as the shape of the mountains or volcanoes and their internal density distributions are highly variable, and since the shower development also depends on the altitude, we calculated the complete directional flux of high-energy secondaries at different altitudes, from the base to the summit every $500$\,m.
As an example, in the figure~\ref{fig:he} we show the expected and normalized flux of high energy secondaries corresponding to $t=1$ year of flux at three different ANDES altitudes: $4450$\,m a.s.l. (near the future tunnel entrance, left panel); $4950$\,m a.s.l. (mid altitude, central panel); and $5450$\,m a.s.l. (mountain summit, right panel).

Muons, charged pions and even nucleons can be observed in the figure~\ref{fig:he} at the TeV/c momentum scale, and they need to be accounted as, e.g., a 100\,TeV charged pion could decay in the rock in a highly penetrating muon.
In future studies, we will select from each one of these libraries only those particles that, according to their altitude and trajectories, are expected to reach the underground lab location or the detector on the other side of the volcano.
As a final remark, it is important to notice that, e.g., a 24-hour flux in one of the high latitude sites involves the simulation of $\sim 1.9\times 10^9$ different showers from protons to irons.
Thus, and having in mind our detailed calculations for ANDES and other underground labs and muography sites, in this first cloud-based test of intensive data production we simulate the impressive figure of $10^{11}$ in more than 130,000 hours-processors for S0 production plus more than 50,000 hours-processors for S1 analysis.
Results generated during these runs are taking up over >3 TB for the S0 bzip2-compressed plus >0.5 TB for the S1 bzip2-compressed synthetic data and metadata.

%% file: author.tex
\clearpage
\section*{Full Authors List: \Coll\ LAGO Collaboration}
\scriptsize
\noindent
V. Agosín$^{17}$,
A. Alberto$^{3}$,
C. Alvarez$^{16}$,
J. Araya$^{20}$,
R. Arceo$^{16}$,
O. Areso$^{13}$,
L. H. Arnaldi$^{32}$,
H. Asorey$^{14,7}$,
M. Audelo$^{9}$,
M.G. Ballina-Escobar$^{19}$
% ...
\newline

\noindent
% ...
$^{3}$CIEMAT.
% ...
$^{7}$Departamento Física Médica, CNEA-CONICET-UNSAM.
% ...
$^{9}$Escuela Superior Politécnica de Chimborazo.
% ...
$^{13}$Instituto de Astronomía y Física del Espacio, IAFE (UBA-CONICET).
$^{14}$Instituto de Tecnologías en Detección y Astropartículas (CNEA, CONICET, UNSAM).
$^{16}$Universidad Autónoma de Chiapas.
$^{17}$Universidad de La Serena.
% ...
$^{19}$Universidad de San Carlos.
$^{20}$Universidad de Valparaíso.
% ...
$^{32}$Centro Atómico Bariloche (CNEA/IB).
% ...
